\documentclass[aps,pra,twocolumn,longbibliography] {revtex4-1}
\usepackage{amsmath,graphicx,amsfonts}

\usepackage{times}
\usepackage[colorlinks=true,linkcolor=blue,urlcolor=blue,citecolor=blue, breaklinks=true,hypertexnames=false]{hyperref}

\usepackage{amssymb}
\usepackage{bbm}

\usepackage[utf8]{inputenc}
\DeclareUnicodeCharacter{03BC}{\mbox{$\mu$}}

\DeclareUnicodeCharacter{03B4}{$\delta$}
\DeclareUnicodeCharacter{2009}{-}

\usepackage[english]{babel}
\addto\captionsenglish{}
\addto\captionsenglish{}

\makeatletter
\def\bbl@set@language#1{%
	\edef\languagename{%
		\ifnum\escapechar=\expandafter`\string#1\@empty
		\else\string#1\@empty\fi}%
	\@ifundefined{babel@language@alias@\languagename}{}{%
		\edef\languagename{\@nameuse{babel@language@alias@\languagename}}%
	}%
	\select@language{\languagename}%
	\expandafter\ifx\csname date\languagename\endcsname\relax\else
	\if@filesw
	\protected@write\@auxout{}{\string\select@language{\languagename}}%
	\bbl@for\bbl@tempa\BabelContentsFiles{%
		\addtocontents{\bbl@tempa}{\xstring\select@language{\languagename}}}%
	\bbl@usehooks{write}{}%
	\fi
	\fi}
\newcommand{\DeclareLanguageAlias}[2]{%
	\global\@namedef{babel@language@alias@#1}{#2}%
}
\makeatother

\DeclareLanguageAlias{en}{english}

\makeatletter
\def\@bibdataout@aps{%
	\immediate\write\@bibdataout{%
		@CONTROL{%
			apsrev41Control%
			\longbibliography@sw{%
				,author="08",editor="1",pages="1",title="0",year="1"%
			}{%
				,author="08",editor="1",pages="1",title="",year="1"%
			}%
		}%
	}%
	\if@filesw \immediate \write \@auxout {\string \citation {apsrev41Control}}\fi 
}
\makeatother

\date\today
\begin{document}
\title{Method of difference-differential equations for some Bethe ansatz solvable models}
\author{Zoran Ristivojevic}
\affiliation{Laboratoire de Physique Th\'{e}orique, Universit\'{e} de Toulouse, CNRS, UPS, 31062 Toulouse, France}
	
\begin{abstract}
In studies of one-dimensional Bethe ansatz solvable models, a Fredholm integral equation of the second kind with a difference kernel on a finite interval often appears. This equation does not generally admit a closed-form solution and hence its analysis is quite complicated. Here we study a family of such equations concentrating on their moments. We find exact relations between the moments in the form of difference-differential equations. The latter results significantly advance the analysis, enabling one to practically determine all the moments from the explicit knowledge of the lowest one. As applications, several examples are considered. First, we study the moments of the quasimomentum distribution in the Lieb-Liniger  model and find explicit analytical results. The latter moments determine several basic quantities, e.g., the $N$-body local correlation functions. We prove the equivalence between different expressions found in the literature for the three-body local correlation functions and find an exact result for the four-body local correlation function in terms of the moments of the quasimomentum distributions. We eventually find the analytical results for the three- and four-body correlation functions in the form of asymptotic series in the regimes of weak and strong interactions. Next, we study the exact form of the low-energy spectrum of a magnon (a polaron) excitation in the two-component Bose gas described by the Yang-Gaudin model. We find its explicit form, which depends on the moments of the quasimomentum distributions of the Lieb-Liniger model. Then, we address a seemingly unrelated problem of capacitance of a circular capacitor and express the exact result for the capacitance in the parametric form. In the most interesting case of short plate separations, the parametric form has a single logarithmic term. This should be contrasted with the explicit result that has a complicated structure of logarithms.
\end{abstract}
\maketitle

\section{Introduction}

Since its discovery in 1931, the Bethe ansatz method has led to the exact solutions for several quantum many-body problems. They describe various physical systems ranging from one-dimensional magnets and quantum impurities interacting with environments to  ultracold quantum gases, which can now be  experimentally realized \cite{guan_fermi_2013,cazalilla_one_2011}. These theoretical achievements have thus taken a central place with a long-lasting impact on modern physics. The exact solutions are without limitations on the physical parameters and thus are of paramount importance, serving as a valuable input for further developments. They are firm grounds for theoretical studies, benchmarks for computer simulations, and challenges for experimental probes.

Arguably one of the simplest many-body problems that admits an exact Bethe ansatz solution consists of one-dimensional bosons with contact interaction. It is known as the Lieb-Liniger model \cite{lieb_exact_1963}. This archetypal example enhances our understanding of quantum physics of interacting particles with numerous theoretical results obtained \cite{korepin1993book,cazalilla_one_2011}. The Lieb-Liniger model has also attracted  significant attention from experimentalists. Early realizations \cite{paredes_tonks-girardeau_2004,kinoshita_observation_2004} were concentrated on the observation of boson fermionization. It occurs in the Tonks–Girardeau regime, where the repulsion is so strong that the bosons behave effectively as free fermions \cite{girardeau_relationship_1960}. This is manifested by the suppression of the local correlation functions. Indeed, the measurements of the two-body \cite{kinoshita_local_2005} and the three-body correlation functions \cite{tolra_observation_2004,haller_three-body_2011} were in agreement with the theoretical predictions \cite{gangardt_stability_2003,cheianov_exact_2006}.

A central quantity that determines various properties of the Lieb-Liniger model is the quasimomentum distribution. In the ground state of the system, it is nonzero between the so-called Fermi quasimomenta. The quasimomentum distribution can be easily found in the case of infinite boson repulsion. Then the quasimomenta coincide with the momenta of a noninteracting spinless Fermi gas. Thus, the quasimomentum distribution has a constant value and the Fermi quasimomenta coincide with the Fermi momenta. By decreasing the repulsion, the quasimomenta evolve according to the Bethe ansatz equations; their distribution begins to shrink symmetrically. At weak repulsion, the distribution width scales with the square root of the interaction strength \cite{lieb_exact_1963}. It thus becomes sharply peaked around zero momentum, marking a tendency of the system toward Bose condensation. The quasimomentum distribution is governed by a linear integral equation \cite{lieb_exact_1963} that does not have a closed-form solution. Nevertheless, the quasimomentum distribution is physically interesting as it has been  directly measured in a recent experiment \cite{wilson_observation_2020}, showing an agreement with the numerical results. 

The quasimomentum distribution determines the ground-state energy via the second moment, which can be routinely evaluated numerically. The analytical results for the ground-state energy in terms of the power series have been obtained in the regimes of weak and strong interactions. Unlike the latter case where a perturbation theory has been developed for the integral equation enabling systematic evaluation of the energy to an arbitrary order \cite{ristivojevic_excitation_2014}, for a long time only the first three terms of the expansion were known at weak interactions  \cite{popov_theory_1977}, despite some controversy \cite{kaminaka_higher_2011,tracy_ground_2016}. Recently, it has been discovered how to obtain more terms in the series expansion \cite{prolhac_ground_2017,ristivojevic_conjectures_2019,marino_exact_2019} and thus practically describe the regime of intermediate interaction strengths analytically. Remarkably, even the spectrum of elementary excitations can be obtained from the quasimomentum distribution of the system in the ground state \cite{petkovic_spectrum_2018}, which emphasizes its importance. The quasimomentum distribution also determines various correlation functions. For example, the short-distance expansion of the one-body density matrix can be expressed as an algebraic expression containing various moments of the distribution \cite{olshanii_short-distance_2003,olshanii_connection_2017}; the exponent of the decay of the one-body density matrix is a function of the value of the distribution at the edge \cite{haldane_effective_1981}.

The $N$-body local correlation functions in the Lieb-Liniger Bose gas have been calculated in several works. In the limiting regimes of weak and strong interactions, explicit results were obtained in Refs.~\cite{gangardt_stability_2003,gangardt_local_2003,kormos_expectation_2009,nandani_higher-order_2016}. In the cases of arbitrary interactions and $N=2$ \cite{gangardt_stability_2003} and $N=3$ \cite{cheianov_three-body_2006,cheianov_exact_2006}, the local correlation functions have been expressed in terms of the moments of the quasimomentum distribution. The case $N=3$ was also exactly solved in Ref.~\cite{kormos_exact_2011}, but the equivalence with the corresponding result of Refs.~\cite{cheianov_three-body_2006,cheianov_exact_2006} was not shown. Finally, the most general case of arbitrary $N$ was solved \cite{pozsgay_local_2011,bastianello_exact_2018} using different techniques. However, the final results of Refs.~\cite{pozsgay_local_2011,bastianello_exact_2018} have significantly more complicated forms, without obvious relations between themselves. Moreover, only numerically they were shown to be consistent with the results for $N=2$ and $N=3$ \cite{gangardt_stability_2003,cheianov_exact_2006}. We eventually note that in Ref.~\cite{pozsgay_local_2011}, the obtained analytical result for the case $N=3$ was shown to be equivalent to that of Ref.~\cite{kormos_exact_2011}.

The above examples illustrate that the existence of exact solutions of integrable models does not imply direct and easy access to the analytical results of particular physical quantities. The main goal of this paper is to develop a formalism that will bridge this gap in the case of some experimentally relevant integrable models. Our formalism enables us to calculate \emph{analytically} various important quantities. Particular attention is devoted to the moments of the quasimomentum distribution, which give rise to direct access to the correlation functions. We derived the exact relations between different moments in the form of a difference-differential equation, easily amenable to the analytical treatment. Another application of the developed formalism is the study of the low-energy magnon spectrum in the Yang-Gaudin model. We obtained the explicit result expressed in terms of the moments of the quasimomentum distribution of the Lieb-Liniger model. As a byproduct, we have established the equivalence between different expressions for the local three-body correlation function found in the literature and obtained the exact result for the four-body one expressed in terms of the moments of the quasimomentum distribution. We have finally addressed the well-known problem of the capacitance of a circular capacitor, which is related to the Lieb-Liniger model \cite{gaudin_boundary_1971}.

This paper is organized as follows. In Sec.~\ref{sec:general} the general formalism is developed that enabled us to treat a class of integral equations common for several Bethe ansatz solvable models. The results of Sec.~\ref{sec:general} are then applied to the Lieb-Liniger model in Sec.~\ref{sec:LL}. An exact difference-differential equation that connects different moments of the quasimomentum distribution is found and solved in the regimes of weak and strong interactions. The obtained results are then used in Sec.~\ref{sec:localcorrelationfunctions} in order to study the local correlation functions. We obtained the exact expression for the four-body correlation function in terms of different moments of the quasimomentum distribution. In Sec.~\ref{sec:magnon} it is shown that the low-energy spectrum of a magnon in the Yang-Gaudin Bose gas is fully determined by the moments of the quasimomentum distribution of the Lieb-Liniger model. In Sec.~\ref{sec:capacitance} we obtained the parametric form of the exact result for the capacitance of a circular capacitor, which is particularly useful at small interplate separations. Finally, in Sec.~\ref{sec:discussion} the obtained results and their implications are discussed. More technical details about the properties of the relevant integral equations are presented in Appendices \ref{appendixA} and {\ref{appendixb}.

\section{General results}\label{sec:general}

Let $\mathcal{F}$ be a linear integral operator that acts on a real function $\rho(k,Q)$ of real variables as 
\begin{align}\label{eq:F}
	\mathcal{F}[\rho(k,Q)]=\rho(k,Q)+\frac{1}{2\pi}\int_{-Q}^{Q} dq\theta'(k-q)\rho(q,Q).
\end{align}
Consider a finite integration limit $Q$ and a kernel $\theta'(k)$ that is an even real function. We want to study the properties of a class of equations
\begin{align}\label{eq:Feq}
	\mathcal{F}[\rho_j(k,Q)]=\frac{k^j}{j!},
\end{align}
where $j\ge 0$ is an integer. Equation (\ref{eq:Feq}) can be classified as a Fredholm integral equation of the second kind with a difference kernel on a finite interval. Without going into the mathematical rigor, we consider continuous  $\theta'(k)$ and assume that Eq.~(\ref{eq:Feq}) admits a unique non-trivial solution that is a differentiable function. The solution $\rho_j(k,Q)$ of Eq.~(\ref{eq:Feq}) is an even function of the first argument for even $j$ and odd for odd $j$. It thus satisfies
\begin{align}\label{eq:parity}
	\rho_j(k,Q)=(-1)^j\rho_j(-k,Q).
\end{align}

Solutions of Eq.~(\ref{eq:Feq}) for different $j$ are not independent. Let us derive some relations among them by applying the derivatives to the operator (\ref{eq:F}) \cite{matveev_effective_2016}. Differentiating Eq.~(\ref{eq:Feq}) with respect to $k$ and performing the partial integration one obtains
\begin{align}\label{eq:Feqderk}
	\mathcal{F}\left[\frac{\partial\rho_j}{\partial k}\right]={}&\frac{\varrho_j}{2\pi}\left[\theta'(k-Q)-(-1)^{j}\theta'(k+Q)\right] +\frac{ k^{j-1}}{(j-1)!}.
\end{align}
Here, we have employed the parity property (\ref{eq:parity}), introduced the abbreviation 
\begin{align}\label{eq:varrho}
\varrho_j(Q)=\rho_j(Q,Q),	
\end{align}
and omitted the explicit dependence on the coordinates. Note that Eq.~(\ref{eq:Feqderk}) also applies for $j=0$. In this case the last term on the right-hand side is zero, which is also formally correct since $1/(-1!)=0$. Similarly, differentiating Eq.~(\ref{eq:Feq}) with respect to $Q$ we obtain
\begin{align}\label{eq:FeqderQ}
	\mathcal{F}\left[\frac{\partial\rho_{j-1}}{\partial Q}\right]={}&-\frac{\varrho_{j-1}}{2\pi} [\theta'(k-Q)-(-1)^{j}\theta'(k+Q)]
\end{align}
for $j\ge 1$. From Eq.~(\ref{eq:FeqderQ}) we directly infer
\begin{align}\label{eq:rhoder1}
\frac{1}{\varrho_{j}(Q)}\frac{\partial\rho_{j}(k,Q)}{\partial Q}=\frac{1}{\varrho_{j+2}(Q)}\frac{\partial\rho_{j+2}(k,Q)}{\partial Q}
\end{align}	
for $j\ge 0$. On the other  hand, a linear combination of Eqs.~(\ref{eq:Feqderk}) and (\ref{eq:FeqderQ}) together with Eq.~(\ref{eq:Feq}) leads to
\begin{align}\label{eq:rhoder2}
	\frac{1}{\varrho_j(Q)}\frac{\partial\rho_{j}(k,Q)}{\partial k}+\frac{1}{\varrho_{j+1}(Q)}\frac{\partial\rho_{j+1}(k,Q)}{\partial Q}=\frac{ \rho_{j-1}(k,Q)}{\varrho_j(Q)},
\end{align}
which applies for $j\ge 0$. In the derivation of Eqs.~(\ref{eq:rhoder1}) and (\ref{eq:rhoder2}) we have used the assumption that $\rho_{-1}(k,Q)=0$ is the only solution of the homogeneous equation $\mathcal{F}[\rho_{-1}(k,Q)]=0$. The latter means that an additional condition on the kernel $\theta'(k)$ might be needed in the most general case. However, it will be fulfilled automatically in our applications, as discussed in Appendix \ref{appendixA}.

\subsection{Moments $A_{j,l}$}

The central quantities of our  interest are the moments of $\rho_{j}$, which we define by
\begin{align}\label{eq:Ajl}
	A_{j,l}(Q)=\frac{1}{2\;\! l!}\int_{-Q}^{Q} dk \rho_j(k,Q) k^l,\quad j,l\ge 0.
\end{align}	 
They obey the symmetry property with respect to the exchange of indices,
\begin{align}\label{eq:AjlAlj}
A_{j,l}=A_{l,j},
\end{align} 
which is shown in Appendix \ref{appendixb}. Due to the parity property (\ref{eq:parity}), $A_{j,l}=0$ for odd $j+l$.

The moments (\ref{eq:Ajl}) are not independent and apart from the symmetry (\ref{eq:AjlAlj}), they satisfy a number of other relations. One of them reads
\begin{subequations}\label{eq:Arelation1}
	\begin{align}
	A_{j,j}+A_{j+1,j-1}={}&\varrho_{j} \varrho_{j+1},\quad j=1,2,\ldots,\\
 A_{0,0}={}&\varrho_{0} \varrho_{1}.
	\end{align}
\end{subequations}
It can be derived as follows. Multiplying Eq.~(\ref{eq:Feqderk}) by $\rho_{j-1}(k,Q)$ and the expression $\mathcal{F}[\rho_{j-1}(k,Q)]=k^{j-1}/(j-1)!$ by $\partial \rho_j/\partial k$, after the integration over $k$ in the interval $-Q<k<Q$, the left-hand sides are identical. From the equality of the right-hand sides one obtains Eq.~(\ref{eq:Arelation1}). Other relations between the moments follow directly from Eqs.~(\ref{eq:rhoder1}) and (\ref{eq:AjlAlj}),
\begin{align}\label{eq:Arelation2}
	\frac{1}{\varrho_l}\frac{\partial A_{j,l}}{\partial Q}= \frac{1}{\varrho_{l+2}}\frac{\partial A_{j,l+2}}{\partial Q},
\end{align}
and Eqs.~(\ref{eq:rhoder2}) and (\ref{eq:AjlAlj}),
\begin{align}\label{eq:Arelation3}
	\frac{1}{\varrho_l}\frac{\partial A_{j,l}}{\partial Q}=\frac{A_{j,l}+A_{j-1,l+1}}{\varrho_{l+1}}.
\end{align}
We notice that the combination of Eqs.~(\ref{eq:Arelation2}) and (\ref{eq:Arelation3}) gives a relation that does not involve the derivatives,
\begin{align}\label{eq:Arelation4}
	\frac{A_{j,l-1}+A_{j-1,l}}{\varrho_l}=\frac{A_{j-1,l+2}+A_{j,l+1}}{\varrho_{l+2}}.
\end{align}
Here a negative index should be understood as $A_{-1,l}=A_{l,-1}=0$, which is consistent with Eq.~(\ref{eq:Ajl}) and $\rho_{-1}=0$.

\subsection{Expressions for $A_{j,l}$ in terms of $\varrho_k$}

Equations (\ref{eq:Arelation1}) and (\ref{eq:Arelation4}) enable us to express the integrals $A_{j,l}(Q)$ defined by Eq.~(\ref{eq:Ajl}) in terms of $\varrho_k(Q)$, see Eq.~(\ref{eq:varrho}). Considering the case $j=0$ we find the relation
\begin{align}\label{eq:A0l}
	A_{0,l}=\varrho_0 \varrho_{l+1},\quad l=0,2,4,\ldots.
\end{align}
Equation (\ref{eq:Arelation4}) for $j=1$ leads to
\begin{align}\label{eq:A1l}
A_{1,l}={}&\varrho_1 \varrho_{l+1}-\varrho_0\varrho_{l+2},\quad l=1,3,5,\ldots.
\end{align}
Equations (\ref{eq:A0l}) and (\ref{eq:A1l}) enable us to reexpress Eq.~(\ref{eq:Arelation4}) in the form
\begin{align}\label{eq:jAlA}
A_{j-1,l}+A_{j,l-1}=\varrho_j \varrho_l,\quad j+l \textrm{ odd}.	
\end{align}
This recurrent equation can be solved \cite{mickens2015difference}. We find
\begin{align}\label{eq:Ajlrr}
A_{j,l}=\sum_{k=0}^{j}(-1)^{j+k}  \varrho_k\, \varrho_{j+l+1-k}	,\quad j+l \textrm{ even}.
\end{align}		
Equation (\ref{eq:Ajlrr}) contains an explicit expression for $A_{j,l}$ defined by Eq.~(\ref{eq:Ajl}) in terms of a sum of pairwise products of $\varrho_j$ functions, see Eq.~(\ref{eq:varrho}). Instead of evaluating the integral of the solution of an integral equation, for some applications it might be advantageous to solve several integral equations and evaluate the solutions at a single point according to Eq.~(\ref{eq:Ajlrr}).

Using Eq.~(\ref{eq:jAlA}), the differential equation (\ref{eq:Arelation3}) becomes
\begin{align}\label{eq:dAdQ}
\frac{\partial A_{j,l}}{\partial Q}=\varrho_j \varrho_l,\quad j+l \textrm{ even}.
\end{align}
Instead of the derivatives with respect to $Q$, it is convenient to change the variables and consider the derivatives with respect to $n=\int_{-Q}^Q dk \rho_0(k,Q)/2\pi$ \footnote{In physical applications $n$ is proportional to the density of particles.}. From the definition (\ref{eq:Ajl}) it then follows $n=A_{0,0}/\pi$ and thus Eq.~(\ref{eq:dAdQ}) gives $\partial n/\partial Q=\varrho_0^2/\pi$. Therefore, Eq.~(\ref{eq:dAdQ}) eventually becomes
\begin{align}\label{eq:dAdn}
	\frac{\varrho_0^2}{\pi}\frac{\partial A_{j,l}}{\partial n}=\varrho_j \varrho_l,\quad j+l \textrm{ even}.
\end{align}

\subsection{Expressions for $A_{j,l}$ in terms of $A_{0,2k}$ and its derivative}

Equations (\ref{eq:A0l}) and (\ref{eq:dAdn}) enable us to write
\begin{align}\label{eq:rodd}
	&\varrho_{2l+1}=\frac{1}{\varrho_0}A_{0,2l},\\
	\label{eq:rodd1}
	&\varrho_{2l}=\frac{\varrho_0}{\pi}\frac{\partial A_{0,2l}}{\partial n},\quad l=0,1,2,\ldots.
\end{align}
\begin{widetext}
Substituting them into Eq.~(\ref{eq:Ajlrr}), we find
\begin{subequations}\label{eq:Ajlsol}
	\begin{gather}\label{eq:Ajlsol(a)}
	A_{2j,2l}=\frac{1}{\pi} \frac{\partial A_{0,2j}}{\partial n} A_{0,2l} +\frac{1}{\pi}\sum_{k=0}^{j-1}\left(\frac{\partial A_{0,2k}}{\partial n} A_{0,2j+2l-2k}-A_{0,2k}\frac{\partial A_{0,2j+2l-2k}}{\partial n}\right),\\ \label{eq:Ajlsol(b)}
	A_{2j+1,2l+1}=\frac{1}{\pi}\sum_{k=0}^{j}\left( A_{0,2k} \frac{\partial A_{0,2j+2l+2-2k}}{\partial n} -\frac{\partial A_{0,2k}}{\partial n} A_{0,2j+2l+2-2k}\right).
\end{gather} 
\end{subequations}	
\end{widetext}
Remarkably, the whole class of integrals (\ref{eq:Ajl}) can be expressed only in terms of $A_{0,2k}$ and its derivative. In other words, the moments of $\rho_0(k,Q)$ determine the moments of all other functions $\rho_j(k,Q)$ defined by Eq.~(\ref{eq:Feq}). Our ultimate goal is therefore to study the even moments of $\rho_0(k,Q)$, i.e., $A_{0,2l}$ since $A_{0,2l+1}=0$ due to the parity.

\subsection{Connection between $A_{0,2l+2}$ and $A_{0,2l}$}

Equation (\ref{eq:Ajlsol(b)}) at $j=0$ becomes
\begin{align}\label{eq:polaron}
	A_{1,2l+1}=n \frac{\partial A_{0,2l+2}}{\partial n}-A_{0,2l+2}.
\end{align}
Acting by the derivative $\partial/\partial n$ to Eq.~(\ref{eq:polaron}) and using Eqs.~(\ref{eq:dAdn}) and (\ref{eq:rodd}), one obtains 
\begin{align}\label{eq:main}
	\frac{\partial^2 A_{0,2l+2}}{\partial n^2}=\pi^2\frac{A_{0,2l}}{\varrho_0^4}.
\end{align} 
Equation (\ref{eq:main}) is another remarkable result. It shows that different moments of $\rho_0$ are actually not independent, but obey the difference-differential equation. In the special case $l=0$, Eq.~(\ref{eq:main}) leads to
\begin{align}\label{eq:aux}
	\frac{\partial^2 A_{0,2}}{\partial n^2}=\frac{\pi^3 n}{\varrho_0^4},
\end{align}
which is a connection between $\varrho_0$  and the second derivative of $A_{0,2}$. One can eventually eliminate $\varrho_0$ from Eq.~(\ref{eq:main}) using Eq.~(\ref{eq:aux}), getting an expression that only involves the moments.

The results of Sec.~\ref{sec:general} are general and go beyond any physical application. They have been derived under the minimal assumptions on the kernel in the integral operator (\ref{eq:F}).

\section{Moments of the quasimomentum distribution of the  Lieb-Liniger model}\label{sec:LL}

In this section we will apply the results of Sec.~\ref{sec:general} to the Lieb-Liniger model \cite{lieb_exact_1963}. It describes one-dimensional nonrelativistic bosons of the mass $m$ that interact via the contact $\delta$-function repulsion of the strength $\hbar^2 c/m$. The latter is encoded into the dimensionless parameter \cite{lieb_exact_1963}
\begin{align}\label{eq:gamma}
	\gamma=\frac{c}{n},
\end{align}
which controls various quantities. In Eq.~(\ref{eq:gamma}), $n$ denotes the density of particles. In the following discussion we implicitly assume the system in the thermodynamic limit.

The Lieb-Liniger model was solved exactly using the Bethe ansatz \cite{lieb_exact_1963,korepin1993book}. Its ground state is characterized by the density of quasimomenta $\rho(k,Q)$ that satisfies the integral equation 
\begin{align}\label{eq:LIE}
	\mathcal{F}[\rho(k,Q)]=\frac{1}{2\pi}.
\end{align}
Here the linear integral operator $\mathcal{F}$ is defined by Eq.~(\ref{eq:F}), which should be supplemented by the kernel
\begin{align}\label{eq:phaseshift}
	\theta'(k)=-\frac{2c}{c^2+k^2},
\end{align}
which follows from the two-body scattering phase shift $	\theta(k)=-2\arctan\left({k}/{c}\right)$. In Appendix \ref{appendixA} we show that the integral equation (\ref{eq:LIE}) for the kernel given by Eq.~(\ref{eq:phaseshift}) obeys necessary conditions in order to apply the formalism of Sec.~\ref{sec:general}. The parameter $Q$ in Eq.~(\ref{eq:F}) is called the Fermi quasimomentum in the physical context. In the ground state, the quasimomenta are between $-Q$ and $Q$. 

The density of quasimomenta $\rho(k,Q)$ determines various quantities. Its integral gives the particle density,
\begin{align}\label{eq:n(Q)}
	n(Q)=\int_{-Q}^{Q} dk \rho(k,Q).
\end{align}
Here we have emphasized that $n$ depends on $Q$. The ground-state energy per particle of the system is given by
\begin{align}\label{eq:eps0}
	\epsilon_0=\frac{\hbar^2}{2m n}\int_{-Q}^{Q} dk k^2\rho(k,Q).
\end{align}
It is convenient to express the non-trivial dependence on the interaction in $\epsilon_0$ is terms of the parameter $\gamma$ of Eq.~(\ref{eq:gamma}). Equation (\ref{eq:eps0}) then acquires the form
\begin{align}
	\epsilon_0=\frac{\hbar^2 n^2}{2m} e_2(\gamma).
\end{align}
Here $e_2(\gamma)$ is the special case $l=1$ of the family of dimensionless functions
\begin{align}\label{eq:e2l}
	e_{2l}(\gamma)=\frac{1}{n^{2l+1}} \int_{-Q}^{Q} dk k^{2l}\rho(k,Q).
\end{align}
The right-hand side of Eq.~(\ref{eq:e2l}) is formally a function of $Q$, while on the left-hand side we wrote the dependence on $\gamma$. This is possible since $Q$ is related to $n$ via Eq.~(\ref{eq:n(Q)}) and thus $Q$ can also be related to $\gamma$ of Eq.~(\ref{eq:gamma}) [see Eq.~(\ref{eq:Qfin}) below].

The family of functions $e_{2l}(\gamma)$ are proportional to the even moments of the quasimomentum distribution. We notice that the odd moments vanish. Omitting the trivial proportionality factor, $e_{2l}(\gamma)$ will be loosely called the moments in the following. They determine a number of physically relevant quantities -- the ground-state energy, the local correlation functions, the short-distance expansion of the one-body density matrix, etc., -- as discussed in the Introduction.

The formalism of Sec.~\ref{sec:general} for the special choice of the kernel (\ref{eq:phaseshift}) and $\rho_0=2\pi \rho$ can be applied to study the Lieb-Liniger model. The family of functions (\ref{eq:e2l}) is related to the quantities defined by Eq.~(\ref{eq:Ajl}) by
\begin{align}\label{eq:e2ldefinition}
	A_{0,2l}=\frac{\pi}{(2l)!}{n^{2l+1}}{e_{2l}(\gamma)}.
\end{align}
Therefore, various identities that we previously derived for $A_{0,2l}$ translate into a new set of identities among $e_{2l}$ functions. Our particular focus will be on Eq.~(\ref{eq:main}) as well the special case of  Eq.~(\ref{eq:dAdQ}), which is
\begin{align}\label{eq:Qn}
	\frac{\partial Q}{\partial n} = \frac{\pi}{\varrho_0^2}.
\end{align}
Here $\varrho_0=2\pi\rho(Q,Q)$. Let us first transform the differentiation with respect to $n$ into the one with respect to $\gamma$, where we should use the rules
\begin{align}\label{eq:derivatives}
	\frac{\partial}{\partial n}=-\frac{\gamma}{n}\frac{\partial}{\partial \gamma}, \quad \frac{\partial^2}{\partial n^2}=\frac{2\gamma}{n^2} \frac{\partial}{\partial \gamma}+\frac{\gamma^2}{n^2}\frac{\partial^2}{\partial \gamma^2}.
\end{align}
From Eq.~(\ref{eq:main}) we then directly obtain
\begin{align}\label{eq:diff-diffeq}
	\gamma^{2l+4}	\frac{d^2}{d\gamma^2} \left(\frac{e_{2l+2}(\gamma)}{\gamma^{2l+2}}\right)= \pi^2(2l+1)(2l+2)\frac{e_{2l}(\gamma)}{\varrho_0^4}.
\end{align}
Equation~(\ref{eq:diff-diffeq})  is a new exact relation between the moments (\ref{eq:e2l}). It has a form of the difference-differential equation.

In the special case $l=0$, Eq.~(\ref{eq:diff-diffeq}) becomes
\begin{align}\label{eq:defK}
\gamma^{4}	\frac{d^2}{d\gamma^2} \left(\frac{e_{2}(\gamma)}{\gamma^{2}}\right)= \frac{2\pi^2}{\varrho_0^4},	
\end{align}
where we have used $e_0=1$ obtained from the definition (\ref{eq:e2l}). At this point it is useful to recall that $\varrho_0$ is in fact related to the so called Luttinger liquid exponent $K$ by the relation \cite{haldane_effective_1981,korepin1993book}
\begin{align}\label{eq:Kll}
	\varrho_0=\sqrt{K}.
\end{align}
On the other hand, the Lieb-Liniger model is Galilean invariant, which implies the relation between the sound velocity $v$ and $K$ of the form $mvK=\pi\hbar n$ \cite{haldane_effective_1981}. Equation (\ref{eq:defK}) then reduces to the  thermodynamic relation \cite{lieb_exact2_1963}
\begin{align}
v=\sqrt{\frac{L}{m n}\frac{\partial^2 E_0}{\partial L^2}},
\end{align} 
which expresses $v$ in terms of the derivative of the ground-state energy $E_0=nL\epsilon_0$ with respect to the system size $L$. Here $nL$ corresponds to the total number of particles.

Eliminating $\varrho_0$ from Eq.~(\ref{eq:diff-diffeq}) using Eq.~(\ref{eq:defK}), we obtain
\begin{align}\label{eq:e2l1}
\frac{d^2}{d\gamma^2} \left(\frac{e_{2l+2}(\gamma)}{\gamma^{2l+2}}\right)= (l+1)(2l+1) \frac{d^2}{d\gamma^2} \left(\frac{e_{2}(\gamma)}{\gamma^{2}}\right) \frac{e_{2l}(\gamma)}{\gamma^{2l}}.
\end{align}
For $l=0$, Eq.~(\ref{eq:e2l1}) reduces to an identity, while for $l>0$ it gives the connections between the consecutive terms of the family (\ref{eq:e2l}). Equation~(\ref{eq:e2l1}) is our starting point for the evaluation of  $e_{2l}(\gamma)$ for $l>1$ using the knowledge of $e_2(\gamma)$, which serves as an initial value of the family $e_{2l}(\gamma)$ that generates $l>1$ terms. Since $e_2(\gamma)$ is analytically known in terms of the power series in the regimes of weak and strong interactions, we will be able to evaluate $e_{2l}(\gamma)$ in the two regimes.

\subsection{Weak interactions}

In the regime of weak interactions, $\gamma\ll 1$, the leading-order solution of Eq.~(\ref{eq:LIE}) is $\rho(k,Q)=\sqrt{Q^2-k^2}/2\pi c$ \cite{lieb_exact_1963}. This yields the order of magnitude estimate for the leading-order term in Eq.~(\ref{eq:e2l}),
\begin{align}
	e_{2l}(\gamma)\sim \frac{1}{\gamma}\left(\frac{Q}{n}\right)^{2l+2}.
\end{align} 
Using $e_0=1$, we find $Q\sim n\sqrt{\gamma}$ and thus $e_{2l}(\gamma)\sim \gamma^l$. Since the subsequent terms in the expansion of $e_2$ are multiplied by  $\sqrt{\gamma}$, we should assume the series
\begin{align}\label{eq:e2lass}
	e_{2l}(\gamma)=\sum_{j=0}^{\infty} a_j^{(2l)}\gamma^{l+j/2},
\end{align}
where the values of the coefficients $a_j^{(2l)}$ for $l>1$ will be calculated using the known values of $a_j^{(2)}$ \cite{marino_exact_2019,ristivojevic_conjectures_2019}. Substitution of the form (\ref{eq:e2lass}) into Eq.~(\ref{eq:e2l1}) yields the connection between the coefficients $a_{k}^{(2l+2)}$ from the left-hand side of Eq.~(\ref{eq:e2l1}) and the ones from the right-hand side,
\begin{align}\label{eq:akoff}
&\left(2l+2-k\right) \left(2l+4-k\right)	a_{k}^{(2l+2)} \notag\\
&=(l+1)(2l+1) \sum_{j=0}^{k} (j-2)(j-4) a_{j}^{(2)} a_{k-j}^{(2l)}.
\end{align}
For $l=0$, Eq.~(\ref{eq:akoff}) becomes trivial since $a_{k-j}^{(0)}=\delta_{k,j}$, while for $l>1$ it enables us to evaluate the coefficients in the series (\ref{eq:e2lass}) for $e_{2l}$ using the ones of $e_2$. 

\begin{table}\label{table}
	\caption{Values of the coefficients in the series (\ref{eq:e2lass}) evaluated from Eqs.~(\ref{eq:a0a1a2a3}) using the known values of $a_k^{(2)}$. \label{table}}
	
	\begin{ruledtabular}
		\begin{tabular}{l|l|l	|l|l}
			$a_k^{(2l)}$ &$k=0$&$k=1$ &$k=2$&$k=3$\\ \hline
			$l=1$ & $1$ & $-\frac{4}{3\pi}$& $\frac{1}{6}-\frac{1}{\pi^2}$ & $-\frac{1}{2\pi^3}+\frac{3\zeta(3)}{8\pi^3}$\\
			$l=2$ & $2$ &$-\frac{88}{15\pi}$ & $1-\frac{2}{\pi^2}$ &$-\frac{4}{3\pi}+\frac{1}{\pi^3}+\frac{21\zeta(3)}{4\pi^3}$ \\
			$l=3$& $5$ & $-\frac{824}{35\pi}$ & $5+\frac{14}{3\pi^2}$ & $-\frac{44}{3\pi}+\frac{17}{\pi^3}+\frac{165\zeta(3)}{4\pi^3}$ \\
			$l=4$ & $14$ & $-\frac{29168}{315\pi}$& $\frac{70}{3}+\frac{3452}{45\pi^2}$ & $-\frac{1648}{15\pi}+\frac{1438}{15\pi^3}+\frac{525\zeta(3)}{2\pi^3}$  \\
		\end{tabular}
	\end{ruledtabular}
\end{table}

For a fixed $k$, Eq.~(\ref{eq:akoff}) can be explicitly solved since it is equivalent to a first-order linear difference equation \cite{mickens2015difference}. Using the known values of $a_j^{(2)}$ given in Table \ref{table}, we obtain
\begin{subequations}\label{eq:a0a1a2a3}
	\begin{align}
		a_{0}^{(2l)}={}&\frac{(2l)!}{l\;\! !(l+1)!},\\
		a_{1}^{(2l)}={}&-\frac{16^l}{2\pi}\frac{(l\;\! !)^2 }{(2l+1)!}\sum_{w=0}^{l-1}\frac{(2w+1)!}{16^w (w\;\! !)^2}a_0^{(2w)},\\
		a_{2}^{(2l)}={}& \frac{(2l)!}{2(l-1)! l\;\! ! } \left[\frac{1}{6}-\frac{1}{\pi^2} -\frac{1}{\pi} \sum_{w=1}^{l-1} \frac{(w-1)! w\;\! ! \;\!  a_{1}^{(2w)}}{(2w)!}\right],\\
		a_{3}^{(2l)}={}& \frac{16^l(2l-1)(l!)^2}{64\pi^3 (2l)!}\notag\\
		&\times\sum_{w=0}^{l-1} \frac{(2w)!\left[(4-3\zeta(3))a_0^{(2w)}-32\pi^2 a_{2}^{(2w)}\right]}{16^w (2w-1)(w\;\! !)^2}.
	\end{align}
\end{subequations}
Equations (\ref{eq:a0a1a2a3}) determine the first four coefficients in Eq.~(\ref{eq:e2lass}) for all the moments $e_{2l}(\gamma)$ of the quasimomentum distribution (\ref{eq:e2l}) in the regime of weak interactions. This remarkable result has its roots in the integrability of the Lieb-Liniger model and is one application of the formalism previously derived in Sec.~\ref{sec:general}. In Table \ref{table} we give the analytical values for  $a_{k}^{(2l)}$ for $1\le l\le 4$. A motivated reader can easily generate the coefficients for higher values of $l$ using Eqs.~(\ref{eq:a0a1a2a3}).

The cases $k=2l+2$ and $k=2l+4$ are special for Eq.~(\ref{eq:akoff}) since the left-hand side then nullifies. Therefore the coefficients $a_{2l+2}^{(2l+2)}$ and $a_{2l+4}^{(2l+2)}$ cannot be immediately recursively expressed though the right-hand side of Eq.~(\ref{eq:akoff}). However, at $k=2l+2$ the right-hand side constitutes a new relation enabling one to express the latter missing coefficient,
\begin{align}\label{eq:akoff1}
	a_{2l+2}^{(2l)} = -\frac{1}{8a_0^{(2)}} \sum_{j=1}^{2l+2} (j-2)(j-4) a_j^{(2)} a_{2l+2-j}^{(2l)}.
\end{align}
For $k=2l+4$, Eq.~(\ref{eq:akoff}) gives the relation
\begin{align}
	\label{eq:constraint2}
	\sum_{j=0}^{2l+4} (j-2)(j-4)a_{j}^{(2)} a_{2l+4-j}^{(2l)}=0.
\end{align}
The sum of Eq.~(\ref{eq:constraint2}) does not involve the coefficients $a_{2l}^{(2l)}$ and $a_{2l+2}^{(2l)}$. However, Eq.~(\ref{eq:constraint2}) is a nontrivial relation among the other coefficients of the two series for $e_2$ and $e_{2l}$. Interestingly, Eqs.~(\ref{eq:akoff1}) and (\ref{eq:constraint2}) at $l=1$  lead to the constraints among the coefficients of $e_2$. This means that even within the same moment not all the coefficients are independent \cite{ristivojevic-unpublished}.

We have not found a way to calculate $a_{2l}^{(2l)}$  from the difference-differential equation (\ref{eq:e2l1}). On the practical side, by increasing $l$ in the series (\ref{eq:e2lass}), $a_{2l}^{(2l)}$ becomes progressively less important since it only determines the $2l$-th correction term of the series representation for $e_{2l}$. Theoretically, one can extend the developed methods for $e_2$ \cite{marino_exact_2019,ristivojevic_conjectures_2019} to the case-by-case study of $e_4$, $e_6$, etc., in order to obtain $a_{2l}^{(2l)}$. We performed this rather involved work. For curious readers we give the final results:
\begin{align}
	a_{4}^{(4)}={}&\frac{3}{20}-\frac{1}{\pi^2}-\frac{21\zeta(3)-10}{6\pi^4},\\
	a_{6}^{(6)}={}&\frac{61}{168}-\frac{9}{4\pi^2}-\frac{5(21\zeta(3)-10)}{12\pi^4}- \frac{3}{5120\pi^6}\bigl(2048\notag\\
	&+15460\zeta(3)-43050\zeta(3)^2+122505\zeta(5)\bigr).
\end{align}
Obviously, the coefficients $a_{2l}^{(2l)}$ become progressively more complicated as $l$ is increased. An interested reader can use the coefficients $a_{4}^{(4)}$ and $a_{6}^{(6)}$ and the ones of $e_2$ found in Ref.~\cite{ristivojevic_conjectures_2019} to easily extend the values listed in Table~\ref{table} to $k\le 7$ for arbitrary $l$ by iterating Eqs.~(\ref{eq:akoff}) and (\ref{eq:akoff1}). On the other hand for $l=2$ and $l=3$ (i.e., for the evaluation of $e_4$ and $e_6$) there is no intrinsic limitation on $k$, the only one being the knowledge of $e_2$.

\subsection{Strong interactions}

In the regime of strong interactions, $\gamma\gg 1$, the integral in the integral operator of Eq.~(\ref{eq:LIE}) is subdominant. This directly leads to $\rho(k,Q)=1/2\pi$ at the leading order, and thus $e_{2l}(\gamma)\sim 1$. Since the subsequent terms in $\rho(k,Q)$ are by a factor of $1/\gamma$ smaller, the resulting series for its moments can be assumed in the form
\begin{align}\label{eq:e2gass}
	e_{2l}(\gamma)=\sum_{j=0}^{\infty} b_j^{(2l)}\gamma^{-j}.
\end{align}
Substituting Eq.~(\ref{eq:e2gass}) into Eq.~(\ref{eq:e2l1}) we find an equation
\begin{align}\label{eq:diffg}
	b_k^{(2l+2)}={}&\frac{(l+1)(2l+1)}{(2l+2+k)(2l+3+k)}\notag\\
	&\times\sum_{j=0}^{k}(2+j)(3+j)b_j^{(2)} b_{k-j}^{(2l)}
\end{align}
that relates the coefficients of Eq.~(\ref{eq:e2gass}). Equation (\ref{eq:diffg}) is a difference equation that has a similar structure as Eq.~(\ref{eq:akoff}), and thus it can be solved for $l>1$. The first five terms are given by
\begin{subequations}
	\label{eeq1}
\begin{align}
	b_0^{(2l)}={}&\frac{\pi^{2l}}{2l+1},\quad b_1^{(2l)}=-\frac{4l\;\!\pi^{2l}}{2l+1},\quad
	b_2^{(2l)}=4l\pi^{2l},\\ 
	b_3^{(2l)}={}&-\frac{16l(l+1)\pi^{2l}}{3}\biggl[1-\frac{\pi^2}{(2l+1)(2l+3)} \biggr],\\
	b_4^{(2l)}={}&\frac{8l(l+1)(2l+3)\pi^{2l}}{3}\biggl[1-\frac{4\pi^2}{(2l+1)(2l+3)}\biggr].\!
	\end{align}
\end{subequations}
Here we have used the known values of $b_j^{(2)}$ entering $e_2$ \cite{ristivojevic_excitation_2014}. They can be recovered from Eqs.~(\ref{eeq1}) setting $l=1$. We note that at strong interactions, the knowledge of $e_2$ suffices to find all other momenta using Eq.~(\ref{eq:diffg}) due to the physical reason of not having divergent moments $e_{2l}$ at $\gamma\to\infty$. This should be contrasted with the regime of weak interactions where in addition to $e_2$ one also needs the ``diagonal'' coefficients $a_{2l}^{(2l)}$ for $l>1$ in order to evaluate $a_{k}^{(2l)}$ at $k\ge 4$.

\subsection{Fermi quasimomentum}

Let us find an expression for the Fermi quasimomentum in terms of $\gamma$.
Its density derivative is given by Eq.~(\ref{eq:Qn}). By making use of Eq.~(\ref{eq:defK}), the Fermi quasimomentum can be expressed as
\begin{align}\label{eq:Qfin}
	Q={}&2n\sqrt\gamma\, g(\gamma),
\end{align}
where $g(\gamma)$ satisfies a differential equation
\begin{align}\label{eq:gdiff}
	\frac{d}{d\gamma}\left(\frac{g(\gamma)}{\sqrt\gamma}\right) = -\sqrt{\frac{1}{8}  \frac{d^2}{d\gamma^2} \left(\frac{e_{2}(\gamma)}{\gamma^{2}}\right)}.
\end{align}
Therefore, the nontrivial dependence in $Q$ is encoded into the latter differential equation, which we solve now.

In the regime of weak interactions, $\gamma\ll 1$, using the result for $e_2$ we find
\begin{align}	\label{eq:gsolutionweak}
	g(\gamma)={}&1-\frac{\sqrt\gamma}{4\pi} \left(\ln\frac{32\pi}{\sqrt\gamma}-1 \right) +\frac{\gamma}{32\pi^2}+\frac{3(\zeta(3)-1)}{256\pi^3}\gamma^{3/2}\notag\\
	&+O(\gamma^2).
\end{align}
The integration constant of the first-order equation (\ref{eq:gdiff}) is the constant term proportional to $\sqrt{\gamma}$ in Eq.~(\ref{eq:gsolutionweak}). Its value is set using the known perturbative solution \cite{popov_theory_1977} of Eq.~(\ref{eq:LIE}) that enables one to find the subleading-order terms in $Q$, which in turn determines the integration constant. We should note that the function of Eq.~(\ref{eq:gsolutionweak}) and thus $Q/n$ has only one logarithmic term unlike the inverse relation where the same logarithm proliferates. The situation is simpler at $\gamma\gg 1$ since the integration constant for Eq.~(\ref{eq:gdiff}) must be set to zero due to the physical reason of not having divergent $Q\propto \gamma$. We find
\begin{align}
	g(\gamma)=\frac{\pi}{2\sqrt\gamma}\Biggl(1-\frac{2}{\gamma} +\frac{4}{\gamma^2}+\frac{\frac{4\pi^2}{3}-8}{\gamma^3} + O\left( {\gamma^{-4}}\right) \Biggr).
\end{align}
Substituting this into Eq.~(\ref{eq:Qfin}) we find $Q$ that is in agreement with the expression found in Ref.~\cite{ristivojevic_excitation_2014}.

\section{Local correlation functions in the Lieb-Liniger model}\label{sec:localcorrelationfunctions}

A local $N$-body correlation function is defined as the ground-state expectation value 
\begin{align}\label{eq:gj}
	g_N(\gamma)=\frac{1}{n^N}\left\langle \Psi^\dagger(x)^N\Psi(x)^N\right\rangle
\end{align}
of the  Bose field operators $\Psi^\dagger$ and $\Psi$, which satisfy the canonical commutation relation $[\Psi(x),\Psi^\dagger(y)]=\delta(x-y)$. The result for the particular case $N=2$ can be easily obtained by applying the Feynman-Hellmann theorem to the Hamiltonian of the Lieb-Liniger model, leading to $g_2(\gamma)=de_2(\gamma)/d\gamma$ \cite{gangardt_stability_2003}. In the case of an arbitrary integer $N$, the exact evaluation of the average value in Eq.~(\ref{eq:gj}) is significantly more difficult \cite{pozsgay_local_2011,bastianello_exact_2018}. The final result of Ref.~\cite{pozsgay_local_2011} is expressed as an integral representation
\begin{align}\label{eq:gNexact}
	g_N={}&\frac{(N!)^2}{(2\pi n)^N}\int_{-Q}^Q dq_1\ldots dq_N \prod_{1\le l<j\le N} \frac{q_j-q_l}{(q_j-q_l)^2+c^2}\notag\\
	&\times \prod_{j=1}^{N}(j-1)!\rho_{j-1}(q_j).
\end{align}
Here $\rho_{j}$ satisfies Eq.~(\ref{eq:Feq}), where the kernel in the integral operator (\ref{eq:F}) is given by Eq.~(\ref{eq:phaseshift}). In the case $N=2$, Eq.~(\ref{eq:gNexact}) reduces to the above-mentioned result \cite{gangardt_stability_2003}, while the actual factorization and evaluation in terms of $\gamma$ is still an involved task. Below we consider the cases $N=3$ and $N=4$.

\subsection{The three-body case}

For $N=3$, we can split the product over the two indices in Eq.~(\ref{eq:gNexact}) into a sum that involves six permutations of $q_1$, $q_2$, and $q_3$, which can then be treated term by term. In this way one can obtain the final expression in the form \cite{pozsgay_local_2011}
\begin{align}\label{eq:g3exact2}
	g_3(\gamma)={}&\frac{12}{\pi n^5\gamma^2}\left(-2A_{3,1}+A_{2,2}+2A_{4,0}\right)\notag\\
	&+\frac{1}{\pi n^3}(2A_{2,0}-A_{1,1})-\frac{2}{\pi^2 n^4\gamma}A_{0,0}A_{1,1},
\end{align}
where $A_{j,l}$ is defined by Eq.~(\ref{eq:Ajl}) \footnote{Equation (\ref{eq:g3exact2}) corresponds to Eq.~(7.10) of Ref.~\cite{pozsgay_local_2011} and to Eq.~(7) of Ref.~\cite{kormos_exact_2011} derived in a complementary way. In Eq.~(\ref{eq:g3exact2}) we omitted the term $A_{1,0}$ that nullifies in the ground state. We notice that the object $\{j,l\}$ used in Ref.~\cite{pozsgay_local_2011} is equal to $j!l! A_{l,j}/\pi$ in our notation.}. Equation (\ref{eq:g3exact2}) is expressed in term of various moments of $\rho_j$ and thus it can be further transformed to a more convenient form that only involves the moments of $\rho_0$. Using Eqs.~(\ref{eq:Ajlsol}) in the expression (\ref{eq:g3exact2}) we obtain
\begin{align}\label{eq:g3exact3}
	g_3(\gamma)={}&\frac{12}{\pi n^5 \gamma^2} \left(-3n\frac{\partial A_{0,4}}{\partial n}+5A_{0,4}+\frac{A_{0,2}}{\pi} \frac{\partial A_{0,2}}{\partial n}\right)\notag\\
	&+\frac{2+\gamma}{\pi\gamma n^3}\left(A_{0,2}-n \frac{\partial A_{0,2}}{\partial n}\right)+\frac{2}{\pi n^3}A_{0,2}.
\end{align}
Taking into account the definition (\ref{eq:e2ldefinition}) of $e_{2l}$ and transforming the derivative to be with respect to $\gamma$ according to Eq.~(\ref{eq:derivatives}), Eq.~(\ref{eq:g3exact3}) becomes 
\begin{align}\label{eq:g3exact1}
	g_3(\gamma)=\frac{3e'_4}{2\gamma}-\frac{5e_4}{\gamma^2} +\left(1+\frac{\gamma}{2}\right)e_2'-\frac{2e_2}{\gamma}-\frac{3e_2e_2'}{\gamma}+\frac{9e_2^2}{\gamma^2},
\end{align}
where $e_{2l}'={d e_{2l}(\gamma)}/{d\gamma}$. Equation (\ref{eq:g3exact1}) coincides with the expression initially found in Refs.~\cite{cheianov_three-body_2006,cheianov_exact_2006} using yet another approach. We have therefore proven that the the exact results (\ref{eq:g3exact2}) and (\ref{eq:g3exact1}) are equivalent, which \textit{a priori} was not obvious at all.

\subsection{The four-body case}

The local correlation function (\ref{eq:gNexact}) in the case $N=4$ can be treated in a way similar to $N=3$. This leads to  \cite{pozsgay_local_2011,Note3}\footnotetext{Equation (\ref{eq:g4pozs}) corresponds to Eq.~(7.12) of Ref.~\cite{pozsgay_local_2011} where we omitted the terms proportional to $A_{j,l}$ with odd $j+l$ since they nullify in the ground state.}
\begin{widetext}
\begin{align}\label{eq:g4pozs}
	g_4(\gamma)={}&\frac{8\gamma}{5\pi n^3}(2A_{2,0}-A_{1,1}) -\frac{16}{5\pi^2 n^4}A_{0,0}A_{1,1} +\frac{24}{\pi n^5\gamma}(A_{2,2}-2A_{3,1}+2A_{4,0}) \notag\\
	&+\frac{48}{5\pi^2 n^6\gamma^2}(2A_{1,1}A_{2,0}-5A_{2,0}^2 +5A_{0,0}A_{2,2}-8A_{0,0}A_{3,1}) 
		+\frac{144}{\pi n^7\gamma^3} (2A_{6,0}-2A_{5,1}+2A_{4,2}-A_{3,3}).
\end{align}
\end{widetext}
The latter expression can be simplified. Applying Eqs.~(\ref{eq:Ajlsol}), then using Eq.~(\ref{eq:e2ldefinition}), and finally  transforming the derivative to be with respect to $\gamma$ according to Eq.~(\ref{eq:derivatives}), we find
\begin{align}\label{eq:g4exact}
	g_4(\gamma)={}&\frac{e_6'}{\gamma^2}-\frac{28e_6}{5\gamma^3}-\left( \frac{10}{\gamma}+\frac{104}{5\gamma^2}+\frac{9e_2'}{\gamma^2}-\frac{12e_2}{\gamma^3}\right)e_4\notag\\&+\left(3+\frac{26}{5\gamma}+\frac{3e_2}{\gamma^2}\right) e_4'+\left(\frac{18}{\gamma}+\frac{168}{5\gamma^2}\right)e_2^2\notag\\
	&+\left(\frac{8\gamma}{5}+\frac{4\gamma^2}{5}-6e_2 -\frac{84e_2}{5\gamma}\right)e_2' -\frac{16 e_2}{5}.
\end{align}
Equation (\ref{eq:g4exact}) is our exact result for the four-body local correlation function (\ref{eq:gj}) taken at $N=4$. It is expressed in terms of the moments of the quasimomentum distributions and their first derivative and in this respect has a similar structure as Eq.~(\ref{eq:g3exact1}).

\subsection{Explicit results for $g_3(\gamma)$ and $g_4(\gamma)$}

The forms (\ref{eq:g3exact1}) and (\ref{eq:g4exact}) are particularly convenient for the analytical evaluation. Using our previously derived results for $e_{2l}$, we obtain
\begin{align}\label{eq:g3small}
	g_3(\gamma)={}&1-\frac{6\sqrt\gamma}{\pi} +\frac{3\gamma}{2}-\left( \frac{3}{\pi}-\frac{25}{4\pi^3}-\frac{69\zeta(3)}{16\pi^3}\right)\gamma^{3/2}\notag\\
	&+O(\gamma^2)
\end{align}
at $\gamma\ll 1$ and
\begin{align}\label{eq:g3large}
	g_3(\gamma)={}&\frac{16\pi^6}{15\gamma^6}\left(1-\frac{16}{\gamma} +\frac{144-\frac{144\pi^2}{35}}{\gamma^2} -\frac{960-\frac{660\pi^2}{7}}{\gamma^3}\right) \notag\\
	&+O\left(\gamma^{-10}\right)
\end{align}
at $\gamma\gg 1$. For the other case we find
\begin{align}\label{eq:g4small}
	g_4(\gamma)={}&1-\frac{12\sqrt\gamma}{\pi} +\left(4+\frac{24}{\pi^2}\right)\gamma \notag\\
	&-\left( \frac{24}{\pi}-\frac{65}{2\pi^3}-\frac{93\zeta(3)}{8\pi^3}\right)\gamma^{3/2}+O(\gamma^2)
\end{align}
at $\gamma\ll 1$ and
\begin{align}\label{eq:g4large}
	g_4(\gamma)={}&\frac{1024\pi^{12}}{2625\gamma^{12}}\left(1-\frac{30}{\gamma}+\frac{480-\frac{160\pi^2}{21}}{\gamma^2} \right)+O\left({\gamma^{-15}} \right)
\end{align}
at $\gamma\gg 1$. It is fascinating to note that in order to calculate the leading-order term in Eq.~(\ref{eq:g4large}) we need to know the 12th subleading term in $e_2(\gamma)$. This was achieved using the systematic procedure developed in Ref.~\cite{ristivojevic_excitation_2014}. We note that only the leading- and the subleading-order terms in $g_3$ and $g_4$ were known before \cite{gangardt_stability_2003,cheianov_exact_2006,gangardt_local_2003,nandani_higher-order_2016}. However, they were obtained using  complementary techniques that can hardly be extended to give to a better accuracy. On the other hand, the exact results (\ref{eq:g3exact1}) and (\ref{eq:g4exact}) together with the method described in Sec.~\ref{sec:LL} establish a way to explicitly evaluate analytically $g_3$ and $g_4$ to a large number of terms in the series, the only limitation being the knowledge of $e_2$.

\section{Low-energy spectrum of a magnon in the Yang-Gaudin Bose gas}\label{sec:magnon}

The developed formalism in Secs.~\ref{sec:general} and \ref{sec:LL} has another application in the study of low-energy spectrum of a spin-wave excitation (magnon) in the one-dimensional Bose gas with two internal states (isospin-$\frac{1}{2}$), described by the Yang-Gaudin model. In first quantization, the corresponding Hamiltonian is identical to that of the Lieb-Liniger model. Due to the $SU(2)$ symmetry of the Hamiltonian, the eigenstates can be characterized by the total isospin. In the sector where it is maximal, the system is fully isospin-polarized and thus described by the Lieb-Liniger model. It supports two branches of elementary excitations  \cite{lieb_exact2_1963}. In the sector with one isospin reversed, there is a third excitation branch that describes a spin-wave \cite{li_exact_2003,fuchs_spin_2005}, which can also be understood as a polaron \cite{ristivojevic_exact_2021}. In this excited state, the momentum of the system is given by \cite{zvonarev_spin_2007,ristivojevic_exact_2021}
\begin{subequations}
	\label{eq:peps}
	\begin{align}\label{eq:p}
		p(Q,\eta)=\frac{\hbar}{2\pi} \int_{-Q}^{Q} dk\:\! \rho_0(k,Q)\left[\pi-\theta(2k-2\eta)\right].
	\end{align}
Here $p$ explicitly depends on the Fermi quasimomentum $Q$ and the spin rapidity $\eta$, and $\rho_0(k,Q)$ satisfies the integral equation (\ref{eq:Feq}) with the kernel given by Eq.~(\ref{eq:phaseshift}). The energy of the magnon corresponding to the momentum (\ref{eq:p}) can be expressed as \cite{ristivojevic_exact_2021}
	\begin{align}\label{eq:eps}
		\mathcal{E}(Q,\eta)=-\frac{\hbar^2}{2\pi m}	\int_{-Q}^{Q} dk\:\! \rho_1(k,Q)\theta(2k-2\eta),
	\end{align}
\end{subequations}
where $\rho_1(k,Q)$ satisfies Eq.~(\ref{eq:Feq}) with the kernel (\ref{eq:phaseshift}). 

At $\eta\to+\infty$, the momentum (\ref{eq:p}) and the energy (\ref{eq:eps}) nullify. In order to access small $p$ and $\mathcal{E}$ we expand $\theta(2k-2\eta)$ at $\eta/c\gg 1$. Accounting for the leading- and subleading-order terms, we obtain
\begin{subequations}
	\label{eq:pE}
	\begin{gather}
		p=\frac{\hbar c\:\!n}{\eta}- \frac{\hbar c^3 n}{12\eta^3}+\frac{2\hbar c}{\pi \eta^3} A_{0,2}+\ldots,\\
		\mathcal{E}=\frac{\hbar^2 c}{\pi m \eta^2}A_{1,1}+\frac{6\hbar^2c}{\pi m \eta^4}A_{1,3}-\frac{\hbar^2 c^3}{4\pi m\eta^4}A_{1,1}+\ldots,
			\end{gather}  
\end{subequations}
where we have used the definition (\ref{eq:Ajl}). Upon elimination of the spin rapidity $\eta$, Eqs.~(\ref{eq:pE}) enable us to express the low-momentum spectrum as
\begin{align}\label{eq:E(p)}
	\mathcal{E}(p)=\frac{p^2}{2m^*}-\frac{\nu\;\! p^4}{24 \hbar^2 n^2 m}+\ldots,\quad p\ll p^*\sim \hbar n \sqrt{\frac{m}{m^* \nu}}.
\end{align}
Here $m^*$ is the mass of magnon excitation that is given by
\begin{align}\label{eq:m*}
	\frac{m}{m^*}=\frac{2A_{1,1}}{\pi  c\:\! n^2},
\end{align}
and $\nu$ controls the subleading-order term in the spectrum. It reads
\begin{align}\label{eq:nu}
	\nu=\frac{2A_{1,1}}{\pi c n^2}+\frac{96 A_{0,2}A_{1,1}}{\pi^2 c^3 n^3} -\frac{144 A_{1,3}}{\pi c^3 n^2}.
\end{align}
Equations (\ref{eq:m*}) and (\ref{eq:nu}) are the exact relations for the Yang-Gaudin model of the Bose gas, valid at arbitrary interaction.

Actual evaluation of $m^*$ and $\nu$ directly follows from our previous results. In particular, the definition (\ref{eq:e2ldefinition}) enables us to express the mass of the magnon excitation (\ref{eq:m*}) as \cite{ristivojevic_exact_2021}
\begin{align}\label{eq:m*LL}
	\frac{m}{m^*}=-\gamma^2 \frac{d}{d\gamma}\left(\frac{e_2(\gamma)}{\gamma^2}\right).
\end{align}
The coefficient $\nu$ is obtained using Eqs.~(\ref{eq:polaron}) and (\ref{eq:e2ldefinition}),
\begin{align}
	\nu=6\gamma^2 \frac{d}{d\gamma}\biggl(\frac{e_4(\gamma)}{\gamma^4}\biggr)-\left(\gamma^2+{24e_2(\gamma)}\right) \frac{d}{d\gamma}\biggl(\frac{e_2(\gamma)}{\gamma^2}\biggr).\!\!
\end{align}
The series expansion of Eq.~(\ref{eq:m*LL}) was discussed in Ref.~\cite{ristivojevic_exact_2021}. For the other coefficient we obtain
\begin{align}\label{eq:nugamma<<1}
	\nu={}&\frac{24}{5\pi\sqrt\gamma}-1+\frac{28}{3\pi^2}+ \left(\frac{2}{3\pi}+\frac{7}{\pi^3}-\frac{45\zeta(3)}{4\pi^3}\right)\sqrt{\gamma}\notag\\
	&+O(\gamma^{3/2})
\end{align}
at $\gamma\ll 1$. We note the absence of a term linear in $\gamma$ in Eq.~(\ref{eq:nugamma<<1}). We also find
\begin{align}\label{eq:nugamma>>1}
	\nu=\frac{2\pi^2}{3\gamma}-\frac{4\pi^2}{\gamma^2}+\frac{16\pi^2 +\frac{8\pi^4}{15}}{\gamma^3}+O(\gamma^{-4})
\end{align}
in the regime $\gamma\gg 1$. We can now evaluate the condition of smallness of momenta $p\ll p^*$ in Eq.~(\ref{eq:E(p)}): at $\gamma\ll1$ we find $p^*\sim\hbar n \gamma^{1/4}$ and at $\gamma\gg 1$ we obtain a less restrictive $p^*\sim\hbar n$. The momentum condition at $\gamma\ll 1$ that does not reach $\hbar n$ signals the existence of a qualitatively new behavior of the magnon dispersion at finite momenta. This is indeed correct since at momenta higher than $\hbar n \sqrt{\gamma}$, the magnon dispersion approaches the dispersion of a type-II excitation in the Lieb-Liniger model \cite{ristivojevic_dispersion_2022}. Notice that the dispersion (\ref{eq:E(p)}) and the one of the type-II excitation overlap in a wide region of momenta between $\hbar n\sqrt{\gamma}$ and $p^*$.

Here we have shown that the quadratic and quartic coefficients of the low-energy spectrum of a magnon in the Yang-Gaudin Bose gas are determined by the momenta of the quasimomentum distribution (\ref{eq:e2l}) in the Lieb-Liniger model. The latter statement is correct beyond the first two coefficients. Indeed, the series expansion of $\theta(2k-2\eta)$ in Eqs.~(\ref{eq:peps}) is a power law in $k$ with the positive powers and thus the expressions (\ref{eq:pE}) will depend on $A_{j,l}$ defined by Eq.~(\ref{eq:Ajl}). They can be transformed to $e_{2l}$ using Eqs.~(\ref{eq:Ajlsol}) and (\ref{eq:e2ldefinition}).

\section{Capacitance of a circular plate capacitor}\label{sec:capacitance}

As a final example where the results of Secs.~\ref{sec:general} and \ref{sec:LL} are applied we consider the problem of capacitance of a circular parallel plate capacitor. The goal of the study is to quantitatively understand the effects of the edge on the capacitance. In the idealized case with parallel plates of unit radius at a separation $\kappa\to 0^+$, the capacitance is $\mathcal{C}=1/4\kappa$, where the effects of the edges are neglected. For a long time the effects of the edges were described by the logarithmic corrections of the form \cite{shaw_circulardisk_1970,chew_microstrip_1982}
\begin{align}
	\mathcal{C}(\kappa)=\frac{1}{4\kappa}+\frac{\ln\frac{16\pi}{\kappa}-1}{4\pi} +\frac{\kappa \left(\ln^2\frac{16\pi}{\kappa}-2\right)}{16\pi^2}+o(\kappa),
\end{align}
until the recent work \cite{reichert_analytical_2020} where the procedure to obtain an arbitrary number of corrections is described, with the explicit form to the order $\kappa^7$. The final expression is a very complicated expression where each power of $\kappa$ contains logarithmic terms of the same and all smaller powers. The presence of such a number of logarithmic terms is mathematically rather inelegant, which practically spoils the numerical evaluation.

The  problem of capacitance is encoded into the Love integral equation \cite{love_electrostatic_1949}
\begin{align}
	\label{eq:Love}
	f(x,\kappa)-\frac{\kappa}{\pi}\int_{-1}^{1} dy \frac{f(y,\kappa)}{\kappa^2+(y-x)^2}=1.
\end{align}
It determines the function $f(x,\kappa)$ that enables one to express the capacitance as
\begin{align}\label{eq:cap}
	\mathcal{C}(\kappa)=\frac{1}{2\pi}\int_{-1}^{1}d x f(x,\kappa).
\end{align}
Equation (\ref{eq:Love}) is very similar to Eq.~(\ref{eq:LIE}) with the kernel (\ref{eq:phaseshift}). Therefore, the capacitance (\ref{eq:cap}) can be defined parametrically via $\gamma$ as $\kappa(\gamma)=\gamma {n}/{Q}$, $\mathcal{C}(\gamma)={n}/{Q}$. Here one should have in mind that $n/Q$ is a function of $\gamma$, see Eq.~(\ref{eq:Qfin}). We thus arrive at the final result
\begin{align}\label{eq:Cparametric}
	\kappa(\gamma)=\frac{\sqrt\gamma}{2\;\!g(\gamma)},\quad \mathcal{C}(\gamma)= \frac{1}{2\sqrt\gamma\, g(\gamma)},
\end{align}
where the function $g(\gamma)$ is controlled by Eq.~(\ref{eq:gdiff}). Equation (\ref{eq:Cparametric}) is the exact parametric solution for the capacitance at arbitrary separations $\kappa$. 

The regime of small separations between the plates, $\kappa\ll 1$, corresponds to $\gamma\ll 1$. One should therefore substitute $g(\gamma)$ of Eq.~(\ref{eq:gsolutionweak}) in the parametric form (\ref{eq:Cparametric}). The obtained result for the capacitance has a significant simplification with respect to the explicit form $\mathcal{C}(\kappa)$ given in Ref.~\cite{reichert_analytical_2020}. Presently there is only one logarithmic term originating from Eq.~(\ref{eq:gsolutionweak}), unlike the explicit form $\mathcal{C}(\kappa)$ where the same term proliferates. The function $g(\gamma)$ can be calculated trivially beyond the terms of Eq.~(\ref{eq:gsolutionweak}) using the result for $e_2(\gamma)$ and the differential equation (\ref{eq:gdiff}). Further corrections will only contain the  power law terms of $\sqrt{\gamma}$, but not any logarithms.

\section{Discussion}\label{sec:discussion}

In this paper we have developed the theory for evaluation of the moments of the quasimomentum distribution for a class of Bethe ansatz solvable models. Their common feature is the governing integral equation where the integral operator has the form of Eq.~(\ref{eq:F}). The general theory of Sec.~\ref{sec:general} has its straightforward application to the Lieb-Liniger model. The moments in this model satisfy the exact difference-differential equation (\ref{eq:e2l1}), which has been  solved analytically in the regimes of weak and strong interactions. 

The moments of the quasimomentum distribution appear in several contexts. Up to a trivial multiplicative prefactor they are the ground-state eigenvalues of the higher order Hamiltonians that represent nontrivial higher-order conservation laws for the Lieb-Liniger model \cite{davies_higher_1990}. Another example is the dispersion of a magnon in the Yang-Gaudin Bose gas studied in Sec.~\ref{sec:magnon}. The moments also determine the local $N$-body correlation functions (\ref{eq:gj}) as first shown for $N=3$ in Ref.~\cite{cheianov_exact_2006} and $N=4$ here, see Eq.~(\ref{eq:g4exact}). One expects this to be true more generally based on the general unevaluated result (\ref{eq:gNexact}) and the results of Sec.~\ref{sec:general}. Explicit results for $N\ge 5$ are not known presently.

Haldane \cite{haldane_effective_1981} noticed that the exponent of the decay of the one-body density matrix in the Lieb-Liniger model can be expressed in terms of the single point value of the denisty of quasimomenta $\rho(Q,Q)=\varrho_{0}(Q)/2\pi$ [cf.~Eq.~(\ref{eq:Kll})]. From our analysis performed in Sec.~\ref{sec:general}, it follows that all the moments of the quasimomentum distribution and their derivatives can be expressed in terms of the related quantities $\varrho_j(Q)$ defined by Eq.~(\ref{eq:varrho}). Indeed, using Eqs.~(\ref{eq:rodd}), (\ref{eq:rodd1}), and (\ref{eq:e2ldefinition}) we obtain
\begin{align}
	e_{2l}(\gamma)={}&\frac{(2l)!\sqrt{K}}{\pi}\frac{\varrho_{2l+1}(Q)}{n^{2l+1}},\\
	\gamma \frac{d 	e_{2l}(\gamma)}{d\gamma}={}&(2l+1)e_{2l}(\gamma)- \frac{(2l)!}{\sqrt{K}}\frac{\varrho_{2l}(Q)}{n^{2l}}.
\end{align}
Here in the right-hand sides one should eventually express $Q$ in terms of $\gamma$ [see Eq.~(\ref{eq:Qfin})], which will cancel the powers of $n$ in the denominators. 

This work opens possibilities to address other problems. For example, the results of Sec.~\ref{sec:general} can be directly applied to the Yang-Gaudin model of fermions, which is left for a future work. The one-body momentum distribution at high momenta behaves as $W(p)={C}/p^4+{C}_1/p^6+\ldots$, where the so-called Tan contact is given by ${C}\sim g_2=de_2(\gamma)/d\gamma$ for the Lieb-Liniger model \cite{olshanii_short-distance_2003}. It would be interesting to understand a possible relation between the subleading term controlled by ${C}_1$ and $g_4$, see Eq.~(\ref{eq:g4exact}). Another direction would be to understand whether and how the results of this paper can be extended in order to describe the system at finite temperatures.

\appendix

\section{Properties of the operator (\ref{eq:F}) for the kernel controlled by Eq.~(\ref{eq:phaseshift})\label{appendixA}}

The integral equation (\ref{eq:Feq}) can be considered as a special case of the equation
\begin{align}\label{eq:IEgeneral}
	\left(\mathcal{I}+\frac{1}{\lambda } \mathcal{K}\right)\rho=f
\end{align}
where $\lambda=1$. In Eq.~(\ref{eq:IEgeneral}) we have suppressed the variables in the arguments of the functions, introduced the parameter $\lambda$ and the operators of the identity $\mathcal{I}$ as well as the nontrivial part of the integral operator $\mathcal{K}$. The existence of the unique and nontrivial solution $\rho$ crucially depends on the spectral properties of the operator $\mathcal{I}+ \mathcal{K}/\lambda$.

For the special choice of the kernel $\theta'(k)$ given by Eq.~(\ref{eq:phaseshift}), Eq.~(\ref{eq:IEgeneral}) in the homogeneous case $f=0$ reduces to the eigenvalue problem
\begin{align}\label{eq:ev}
\frac{c}{\pi} \int_{-Q}^{Q} dq \frac{\rho(q,Q)}{c^2+(k-q)^2}=	\lambda \rho(k,Q).
\end{align}
In the limit $c\to 0^+$, under the integral we have a representation of the Dirac $\delta$-function. Therefore, $\lambda=1$ is an eigenvalue at $c\to 0^+$. In the opposite regime $c\gg Q$, the left-hand side of Eq.~(\ref{eq:ev}) is proportional to $1/c$ for normalizable eigenfunctions that we impose. One thus expects $\lambda\sim 1/c$ and the spectrum that satisfies
\begin{align}\label{eq:spectrum}
0<\lambda<1.
\end{align}		
Careful treatment of the eigenvalue problem  (\ref{eq:ev}) shows that the spectrum is nondegenerate and obeys $0<\lambda\le 1-2\arctan(c/Q)/\pi$ \cite{baratchart_solution_2019}. Therefore, we can conclude that at finite positive $c$ and at $Q>0$, the spectrum of the eigenvalue problem (\ref{eq:ev}) satisfies the condition (\ref{eq:spectrum}).

This consideration shows that for the special value $\lambda=1$, which is of our interest in the paper, Eq.~(\ref{eq:ev}) has only a trivial solution $\rho=0$. The Fredholm alternative theorem \cite{book-integralequations} then guarantees that Eq.~(\ref{eq:IEgeneral}) has a unique solution that can be formally expressed as
\begin{align}\label{eq:solrho}
	\rho=\left(\mathcal{I}+\mathcal{K}\right)^{-1} f.
\end{align}	
Here the inverse of the operator is defined by the infinite power series, which is convergent due to the condition (\ref{eq:spectrum}). However, the convergence is very slow at small $c/Q$ \cite{love_electrostatic_1949}, which makes the analytical treatment of the Lieb-Liniger model at weak interactions generally troublesome. For smooth $f$ as in Eq.~(\ref{eq:Feq}), the solution of the integral equation will be a differentiable function. From Eq.~(\ref{eq:solrho}), this can be understood as an infinite sum where each term is differentiable.

\section{A property of the pair of integral equations}\label{appendixb}

Consider a pair of integral equations
\begin{gather}\label{b1}
	\mathcal{F}[\sigma(k,Q)]=g(k),\\
	\label{b2}
	\mathcal{F}[\tau(k,Q)]=h(k),
\end{gather}
where the integral operator $\mathcal{F}$ is defined by Eq.~(\ref{eq:F}) and $g(k)$ and $h(k)$ are arbitrary functions that satisfy minimal requirements (i) there are unique solutions $\sigma(k,Q)$ and $\tau(k,Q)$ and (ii) the solutions satisfy 
\begin{align}\label{b3}
	\int_{-Q}^{Q}dk \int_{-Q}^{Q}dq \sigma(k,Q)\theta'(k-q)\tau(k,Q)=\notag\\	\int_{-Q}^{Q}dq \int_{-Q}^{Q}dk \sigma(k,Q)\theta'(k-q)\tau(k,Q),
\end{align}
with $\theta'(k)=\theta'(-k)$. For example, for $g(k)$ and $h(k)$ in the form of polynomials, the assumptions will be satisfied. Then we have the relation
\begin{align}\label{b4}
	\int_{-Q}^{Q}dk \sigma(k,Q)h(k)=\int_{-Q}^{Q}dk \tau(k,Q)g(k).	
\end{align}
Equation (\ref{b4}) can be directly showed by multiplying Eqs.~(\ref{b1}) and (\ref{b2}), respectively, by $\tau(k,Q)$ and  $\sigma(k,Q)$. After performing the integration over $k$ in the interval $-Q<k<Q$, and using the assumption (\ref{b3}) one obtains identical left-hand sides of the two equations. The right-hand sides then give the property (\ref{b4}). Equation (\ref{eq:AjlAlj}) of the main text directly follows from the property (\ref{b4}) for the choice $g(k)=k^j/j!$, and $h(k)=k^l/l!$, and thus $\sigma(k,Q)=\rho_j(k,Q)$, $\tau(k,Q)=\rho_l(k,Q)$, see Eq.~(\ref{eq:Feq}).


%

\end{document}